\documentclass[%
 reprint,
 amsmath,amssymb,
 aps,
]{revtex4-1}

\usepackage{graphicx}% Include figure files
\usepackage{dcolumn}% Align table columns on decimal point
\usepackage{bm}% bold math
\begin{document}

%\preprint{SUESHEP-TH-1101}

\title{Composite Leptons and Quarks from Hexad Preons}
% Force line breaks with \\

\author{Shun-Zhi Wang~$^{1,2}$}
\email{wsz08@pku.edu.cn}
 %\altaffiliation[Also at ]{Physics Department, XYZ University.}
\affiliation{%
 $^{1}$College of Fundamental Studies,\\
 Shanghai University of Engineering Science,
 Shanghai, 201620, P.R.China\\
 $^{2}$Center for High Energy Physics, Peking University,
 Beijing, 100871,  P.R.China
}%

%\date{\today}

\begin{abstract}
A Hexad Preon model where leptons, quarks and W Z bosons are composite is proposed. Six Hexad Preons transform under $U(3)\otimes U(3)$ local gauge group which is identified with  $U(1)_Q\otimes SU(3)_C\otimes SU(3)_f\otimes U(1)_w$.  All salient features of the standard model can be obtained from the compositeness of leptons and quarks: There are exactly six quarks and six leptons with evident three families (generations); All quantum numbers of leptons and quarks can be given out of that of preons; QED and QCD are given by electro-strong interaction $U(1)_Q\otimes SU(3)_C$ ;  The weak interaction is residual "Van der Waals" forces between preons and dipreons. It is shown that all processes in standard model are just reshuffle of preons. In addition, a possible dark matter candidate is presented. Other questions like the electroweak symmetry breaking, the spin of fermions, the origin of quark and lepton mixing, \textit{etc.}, are also addressed.
\begin{description}
\item[PACS numbers]
%12.10.Dm%Unified theories and models of strong and electroweak interactions
%02.40.Tt,%Complex manifolds
%04.50.-h,%Higher-dimensional gravity and other theories of gravity
% 11.10.Kk,%Field theories in dimensions other than four
 12.10.-g%Unified field theories and models,
 12.60.Rc %Composite models
\end{description}
\end{abstract}


\maketitle

%\tableofcontents

\section{\label{sec:intro}Introduction}

Up to now, twelve elementary particles, six quarks and six leptons , and their antiparticles have been found experimentally. The strong, weak and electromagnetic forces among these particles can be described by the standard model of particle physics to very high accuracy\cite{PDG}. However, the important Higgs boson in the standard model has not been found yet, the mechanism for electroweak symmetry breaking, fermion mass generation and fermion mixing, CP violation \textit{etc}. remain unresolved.

The obvious replication of quark-lepton families and the
hierarchy of their masses and mixings have led to the conjecture that
the ultimate explanations to the problems in the standard model  may be the compositeness of  quarks and leptons\cite{D'Souza:1992tg}.  Usually it is assumed that quarks and leptons (even the W, Z, \textit{etc.}) are composites of more fundamental entities, preons. Preons may be fermions or bosons. A new kind of strong interactions called hypercolor should exist to bind preons together. Leptons, quarks \textit{etc.} are singlets under the  hypercolor  gauge symmetry\cite{Harari:1984us}.

One of the simplest preon model is Harari-Shupe model\cite{Harari:1979,Shupe:1979}, which has two
elementary spin-1/2 preons. The preons carry both hypercolor and
color charges.  One of the preons carries electric
charge of $e/3$, while the other one is neutral.
The two types of preons belong to
the fundamental representations of the unbroken local
gauge group $SU(3)_H\otimes SU(3)_C\otimes U(1)_Q$.
Quarks and leptons are composed of three preons so that they are hypercolor
singlets. Within this framework there is no
global SU(2) isospin symmetry on the preon level.
A important feature of the Harari-Shupe model is the connection between fractional electric charges and colored fermions. However, this model can accommodate only the first generation of quarks and leptons, with the other two generations treated as excitation states of the first generation fermions.

In order to contain six quarks and six leptons, a trinity of preons is proposed\cite{Dugne:2002fq}. In this model, there are three spin-1/2 preons with electric charges $+e/3$, $-2e/3$ and $+e/3$, respectively. Pairs of preons may be tightly bounded by spin-dependent forces into spin-0 "dipreon"s. Each preon and its "supersymmetric" partner, which is the anti-dipreon formed by the other two antipreons, are of identical charges.
Preons (and anti-dipreons) belong to complex representations $\bar{3}$
of the $SU(3)_C$, local gauge group of the ordinary QCD.
The hypercolor dynamics was not specified. The preons are assumed to be
stable and there is a global preon-flavour SU(3) symmetry which is
similar to that of the original quark model. The leptons are bound
states of one preon and a dipreon so that they are color singlet while
the quarks are composed of a preon and an anti-dipreon. It should be noted that quarks here can not be overall singlet of
any hypercolor dynamics since the preon and its "supersymmetric"
partner have the same charges.
%Heavy vector bosons are color singlets formed by preon anti-preon pairs.
Preon trinity can give some understanding of such questions in the standard model like lepton number conservation, the Cabibbo-Kobayashi-Maskawa (CKM) mixings, oscillations and decays between some neutrinos \textit{etc} . However, there is no concept of family or generation within this model. In addition, it predicts many exotic composite states apart the ordinary leptons, quarks and W, Z bosons.

The common difficulty in preon model building comes from the fact that the  masses of leptons and quarks are much smaller than the scale of compositeness. Indeed, the small radius of the quarks and leptons indicates that the compositeness scale must be greater than a few TeV. The disappearance of fermion masses is generally guaranteed by the
existence of chiral symmetries. In order to keep preons as well as quark and lepton bound states massless, the chiral symmetries respected by the
strong preon dynamics must remain unbroken as the composites form. As 't Hooft pointed out in a classic paper\cite{tHooft1980} that in order to preserve above chiral symmetries composite fermions must yield the same chiral anomalies as those appearing in the underlying preon theory. It is
this anomaly matching condition that makes most of existing models
rather complex and cumbersome.

In some cases, the 't Hooft anomaly constraint can be avoided. For example, it is shown that in some higher dimensional supersymmetric unified gauge theories, proper Sherk-Schwarz compactification makes chiral multiplets of composite quarks and leptons massless in four dimensions with all unwanted states (residing in the bulk) being still massive\cite{Chaichian:2001fs}. Since we know nothing exactly about the nature of space-time at high energy, we may in general owe the masslessness of leptons and quarks to the potential energy of the superstrong forces which bind the preons together.

In this paper, we follow the line of Harari-Shupe model and preon trinity, putting the emphasis on the inner symmetries of preons.
We propose that there are six anticommutative fields called Hexad
Preons. The underlying preon dynamics is $U(3)\otimes U(3)$ local gauge theory which may be identified with  $U(1)_Q\otimes SU(3)_C\otimes SU(3)_f\otimes U(1)_w$. We will show that all salient features of the standard model can be understood from the compositeness of leptons and quarks. Most of the defects of Harari-Shupe model and preon trinity mentioned above can be remedied within this Hexad Preon model. In addition, we also offer some new point of view to such questions like the electroweak symmetry breaking, the spin of preons, the nature of weak interaction and the origin of fermion mixing.

\section{\label{sec:Preons}Hexad Preons}

Hexad Preon model consists of six anticommutative fields
\begin{eqnarray}
  \chi_a,~\phi_b,~~~~~~~~~~~(a, b=1,2,3)
\end{eqnarray}
and their complex conjugates (antipreons).
They satisfy the following anticommutative relations
\begin{eqnarray}
&\chi_a^2=\phi_a^2=0,
&\chi_a\chi_b+\chi_b\chi_a=0,\label{anti1}\\
&\phi_a\phi_b+\phi_b\phi_a=0,\
&\chi_a\phi_b+\phi_b\chi_a=0\label{anti2}
 \end{eqnarray}
($a, b=1,2,3$) as well as their complex conjugate relations.

 The underlying preon dynamics is $U(3)\otimes U(3)$ local gauge theory. One $U(3)$ stands for electro-strong interaction while the other is responsible for the weak sector. In other word, the gauge group $U(3)\otimes U(3)$ may be identified with  $U(1)_Q\otimes SU(3)_C\otimes SU(3)_f\otimes U(1)_w$. Here Q is electric charge, $SU(3)_C$ is local gauge group of the ordinary QCD,  $SU(3)_f$ is local preon-flavour symmetry, $w$ is a new weak charg. Therefore in Hexad Preon model electromagnetism is intimately related with strong interaction and has no link to weak interaction. In deed, phenomenologically the laws electromagnetism follow are almost the same as that of strong interaction in contrast to that of weak interaction. Therefore there is no problem of electroweak symmetry braking here.

Hexad preons fall into two groups: $\chi_a~(a=1,2,3)$ and $\phi_a~(a=1,2,3)$. $\chi_a~(a=1,2,3)$ belong to
the fundamental representations of the underlying exact electro-strong
gauge group $U(1)_Q\otimes SU(3)_C$ with electric charge $e/3$. $\phi_a~(a=1,2,3)$ belong to
the fundamental representations of the underlying exact weak
gauge group $SU(3)_f\otimes U(1)_w$ with a new weak charge $1/3$. The preon quantum numbers are summarized in TABLE \ref{tab:table1}. The antipreons
have the same quantum numbers with opposite sign.
\begin{table}[h]%The best place to locate the table environment is directly
                %after its first reference in text
\caption{\label{tab:table1}%
            The quantum numbers of Hexad Preons.}
\begin{ruledtabular}
\begin{tabular}{ccccccc}
&\multicolumn{2}{r}{Rep. of} &\multicolumn{2}{r}{Rep. of}\\
 \textrm{Preons}&\textrm{Q}&$\textrm{SU(3)}_C$&$(T_3,Y)_C$
&$\textrm{SU(3)}_f$&$(T_3,Y)_f$ &$w$\\ \hline
\colrule
$\chi_1$ & $1/3$ &   &  $(1/2,1/3)$& &\\
$\chi_2$ & $1/3$ & $3$  & $(-1/2,1/3)$& &\\
$\chi_3$ & $1/3$ &   & $(0,-2/3)$& &\\
\hline
$\phi_1$ & 0 & &  & & $(1/2,1/3)$&   $1/3$\\
$\phi_2$ & 0 & &  & $3$& $(-1/2,1/3)$&  $1/3$ \\
$\phi_3$ & 0 & &  & & $(0,-2/3)$&   $1/3$\\
\end{tabular}
\end{ruledtabular}
\end{table}

The preons are assumed to be stable. They can be produced or destroyed only in preon-antipreon pairs. So apart from the above local charges in TABLE \ref{tab:table1}, each $\phi$ preon and anti-dipreon is assigned a flavor number $\mathcal{P}$ as follows:
\begin{eqnarray}
 &(\bar{\phi}_2 \bar{\phi}_3),~~
 (\bar{\phi}_3 \bar{\phi}_1),~~
 (\bar{\phi}_1 \bar{\phi}_2): ~~\mathcal{P}=w+1=\frac{1}{3};\\
 &\phi_1,~~ \phi_2,~~ \phi_3: ~~\mathcal{P}=w=\frac{1}{3}.
\end{eqnarray}

In the following, we will show that there are just six quarks and six leptons with evident three families (generations). All quantum numbers of leptons and quarks can be given out of that of preons.

Neutrinos are bound states of two $\phi$ antipreons, i.e. anti-dipreons.
\begin{eqnarray}
 \nu_e:~~(\bar{\phi}_2 \bar{\phi}_3),~~~
 \nu_\mu:~~(\bar{\phi}_3 \bar{\phi}_1),~~~
 \nu_\tau:~~(\bar{\phi}_1 \bar{\phi}_2).\label{neutrino}
\end{eqnarray}
Because of the anticommutative relations (\ref{anti2}), there are exactly three families of neutrinos. Since there is no component of $\chi$ preons,  neutrinos are electric neutral and can only take part in weak interactions.

Each charged lepton consists of three $\chi$ antipreons and one $\phi$ preon.
\begin{eqnarray*}
e^-:(\bar{\chi}_1\bar{\chi}_2\bar{\chi}_3) \phi_1,~
 \mu^-:(\bar{\chi}_1\bar{\chi}_2\bar{\chi}_3) \phi_2,~
 \tau^-:(\bar{\chi}_1\bar{\chi}_2\bar{\chi}_3) \phi_3.
\end{eqnarray*}
Leptons are color singlets but have one unit electric charge. Therefore leptons can take part in both electromagnetic and weak interactions.

The lepton numbers are given as follows:
\begin{eqnarray}
 &&L_e=T_{3f}+\frac{Y_f}{2}+\mathcal{P},\\
 &&L_\mu=-T_{3f}+\frac{Y_f}{2}+\mathcal{P},\\
 &&L_\tau=-Y_f+\mathcal{P}.
\end{eqnarray}
It is obvious that lepton number conservation is the consequence of $SU(3)_f\otimes U(1)_w$ symmetry.

The weak isospin for all leptons can be given by
\begin{eqnarray}
 T_{3w}=-\frac{L}{6}-w.
\end{eqnarray}
where $L$ is lepton numbers for leptons.

The quantum numbers of leptons are summarized in TABLE \ref{tab:table2}.
\begin{table}[htb]%The best place to locate the table environment is directly after its first reference in text
\caption{\label{tab:table2}%
         The preon components and quantum numbers of leptons.}
\begin{ruledtabular}
\begin{tabular}{lccrccc}
\textrm{Leptons}&
\textrm{components}&
\textrm{Q}&
\multicolumn{1}{c} {$L_e$}%\footnote{$L_e=T_{3f}+Y_f/2+w+1$.}}
                   &{$L_\mu$}%\footnote{$L_\mu=-T_{3f}+Y_f/2+w+1$.}}
                   &{$L_\tau$}%\footnote{$L_\tau=-Y_f+w+1$.}}
                   &{$T_3w$}\\%\footnote{$T_{3w}=-L/6-w$.}\\
\colrule\vspace{2mm}
$\nu_e$ & $(\bar{\bar{\phi}}_2 \bar{\bar{\phi}}_3)$ & 0 & 1 &0 &0 & 1/2 \\
$e^-$ & $(\bar{\chi}_1  \bar{\chi}_2 \bar{\chi}_3 )\phi_1$& -1 & 1 &0 &0 & -1/2\\
\hline\vspace{2mm}
$\nu_\mu$ & $(\bar{\bar{\phi}}_3 \bar{\bar{\phi}}_1)$ & 0 & 0 &1 &0 & 1/2\\
$\mu^-$ & $(\bar{\chi}_1  \bar{\chi}_2 \bar{\chi}_3 )\phi_2$& -1 & 0 &1 &0 & -1/2\\
\hline\vspace{2mm}
$\nu_\tau$ & $(\bar{\bar{\phi}}_1 \bar{\bar{\phi}}_2)$ & 0 & 0 &0 &1 & 1/2 \\
$\tau^-$ & $(\bar{\chi}_1  \bar{\chi}_2 \bar{\chi}_3 )\phi_3$& -1 & 0 &0 &1 & -1/2\\
\end{tabular}
\end{ruledtabular}
\end{table}

Quarks contain both $\chi$ and $\phi$ preons. Three up quarks are formed by two $\chi$ preons and two $\phi$ antipreons.
\begin{eqnarray*}
   &&u:(\chi_2 \chi_3)(\bar{\phi}_2 \bar{\phi}_3),(\chi_3 \chi_1)(\bar{\phi}_2 \bar{\phi}_3),(\chi_1 \chi_2))(\bar{\phi}_2 \bar{\phi}_3);\\
   && c:(\chi_2 \chi_3)(\bar{\phi}_3 \bar{\phi}_1),(\chi_3 \chi_1)(\bar{\phi}_3 \bar{\phi}_1),(\chi_1 \chi_2))(\bar{\phi}_3 \bar{\phi}_1);\\
   && t:(\chi_2 \chi_3)(\bar{\phi}_1 \bar{\phi}_2),(\chi_3 \chi_1)(\bar{\phi}_1 \bar{\phi}_2),(\chi_1 \chi_2))(\bar{\phi}_1 \bar{\phi}_2).
\end{eqnarray*}
Up quarks transform as color triplet with 2/3 unit electric charge.

Three down quarks are composed of one $\chi$ antipreon and one $\phi$ preon.
\begin{eqnarray}
   &&d:\bar{\chi}_1\phi_1 ,\bar{\chi}_2\phi_1,\bar{\chi}_3\phi_1;\nonumber\\
   && s:\bar{\chi}_1\phi_2 ,\bar{\chi}_2\phi_2,\bar{\chi}_3\phi_2;\label{dquarks}\\
   && b:\bar{\chi}_1\phi_3 ,\bar{\chi}_2\phi_3,\bar{\chi}_3\phi_3.\nonumber
\end{eqnarray}
Down quarks also carry color charge but have -1/3 unit electric charge. Therefore quarks can take part in all three kind of interactions.

The baryon numbers of quarks are just the preon flavor number:
\begin{eqnarray}
 &&\mathcal{B}=\mathcal{P}=\frac{1}{3}.
\end{eqnarray}

During the processes of strong interaction, $\phi$  preons (dipreons) can be produced only in preon-antipreon (dipreon-antidipreon) pairs. The quark numbers which are conserved only in strong interactions like strangeness (S) \textit{etc}. may be given as in the standard model since they just reflect the fact of preon number conservation.

The weak isospin for all quarks can be given by
\begin{eqnarray}
T_{3w}=-\frac{\mathcal{B}}{2}-w.
\end{eqnarray}
where $\mathcal{B}$ is baryon numbers for quarks.

The quantum numbers of quarks are summarized in TABLE \ref{tab:table3}.

\begin{table}[ht]%The best place to locate the table environment is directly after its first reference in text
\caption{\label{tab:table3}%
       The preon components and quantum numbers of quarks.}
\begin{ruledtabular}
\begin{tabular}{lcccc}
\textrm{quarks}&
\textrm{components}&
\textrm{Q}&
\multicolumn{1}{c} {$\mathcal{B}$}
                   &{$T_{3w}$}\\
\colrule\vspace{2mm}
$u$ & $((\chi_2 \chi_3),(\chi_3 \chi_1),(\chi_1 \chi_2))(\bar{\bar{\phi}}_2 \bar{\bar{\phi}}_3)$ & 2/3 & 1/3 & 1/2 \\
$d$& $(\bar{\chi}_1,\bar{\chi}_2, \bar{\chi}_3 )\phi_1$& -1/3 & 1/3 & -1/2  \\
\hline\vspace{2mm}
$c$ & $((\chi_2 \chi_3),(\chi_3 \chi_1),(\chi_1 \chi_2))(\bar{\bar{\phi}}_3 \bar{\bar{\phi}}_1)$ &2/3 & 1/3  & 1/2 \\
$s$ & $(\bar{\chi}_1,\bar{\chi}_2, \bar{\chi}_3  )\phi_2$& -1/3 & 1/3 & -1/2  \\
\hline\vspace{2mm}
$t$ & $((\chi_2 \chi_3),(\chi_3 \chi_1),(\chi_1 \chi_2))(\bar{\bar{\phi}}_1 \bar{\bar{\phi}}_2)$ &2/3 & 1/3   & 1/2 \\
$b$ & $(\bar{\chi}_1,\bar{\chi}_2, \bar{\chi}_3  )\phi_3$& -1/3 & 1/3 & -1/2\\
\end{tabular}
\end{ruledtabular}
\end{table}

 According to above patterns , it is reasonable that a new composite state, $(\chi_1\chi_2\chi_3)(\bar{\phi}_1\bar{\phi}_2\bar{\phi}_3)$, should exist.  This state is $SU(3)_C$ and $SU(3)_f$ singlet with overall $U(1)$ charge neutral. it is a possible dark matter candidate.

The spin of preons is related to the space-time at the scale of compositeness and will not be discussed here. The leptons and quarks are composed of two or four preons and (or) aitipreons. It is this point that makes Hexad Preon model economical and elegant. In order to understand the spin of leptons and quarks, let's consider the situation in quantum field theory where Weyl and Dirac spinors are composed of two and four components. The spin of spinors is determined meaningfully there and we have never worried about the spin of their components. Perhaps the status of our preons here is similar to that of  spionor components rather than fermion or boson fields. This may be an indication that the nature of space-time at the scale of compositeness should be much different from that of ordinary four dimensional space-time continuum.

\section{The weak interaction and fermion mixing}

In the Hexad Preon model, electro-strong interaction $U(1)_Q\otimes SU(3)_C$ gives the same dynamics as that of ordinary QED and QCD. However, the  gauge interaction of weak sector $SU(3)_f\otimes U(1)_w$ is obviously broken at low energy. This could be realized by the condensate of $\phi$ preons. Pairs of $\phi$ preons  may be tightly bounded into "dipreon"s. The weak interactions are just residual "Van der Waals" forces between preons and dipreons.

Given above preon assignment of quarks and leptons, all processes in standard model are just reshuffle of preons. For example, muon decay
\begin{equation}
\mu^- \rightarrow \nu_\mu+e^-+\bar{\nu_e}  \label{eq:mynum}
\end{equation}
 is as follows:
\begin{equation}
[(\bar{\chi}_1  \bar{\chi}_2 \bar{\chi}_3 )\phi_2] \rightarrow
[(\bar{\phi}_3 \bar{\phi}_1)]
+[(\bar{\chi}_1  \bar{\chi}_2 \bar{\chi}_3 )\phi_1]
+[(\phi_2 \phi_3)] ~. \label{eq:mynum}
\end{equation}

In the case of neutrino lepton interaction, e. g.
\begin{equation}
\nu_\mu+e^-\rightarrow \mu^- +\nu_e, \label{eq:mynum}
\end{equation}
we have
\begin{equation}
[(\bar{\phi}_3 \bar{\phi}_1)]
+[(\bar{\chi}_1  \bar{\chi}_2 \bar{\chi}_3 )\phi_1]\rightarrow
[(\bar{\chi}_1  \bar{\chi}_2 \bar{\chi}_3 )\phi_2]
+[(\bar{\phi}_2 \bar{\phi}_3)] ~. \label{eq:mynum}
\end{equation}

The fact that weak interactions involving quarks have the same patten as that of leptons can be understood now at the preon level. Nuclear beta decay
\begin{equation}
d \rightarrow u+e^-+\bar{\nu}_e ~ \label{eq:mynum}
\end{equation}
is just the process
\begin{equation}
[\bar{\chi}_1\phi_1] \rightarrow
 [(\chi_2 \chi_3 )(\bar{\phi}_2 \bar{\phi}_3)]
+[(\bar{\chi}_1  \bar{\chi}_2 \bar{\chi}_3 )\phi_1]
+[(\phi_2 \phi_3)] ~. \label{eq:mynum}
\end{equation}
Here is another example, Pion decay:
\begin{eqnarray}
&\pi^-(\bar{u}d)\rightarrow \mu^- + \bar{\nu}_\mu~, \label{eq:mynum}\\
&[(\bar{\chi}_2  \bar{\chi}_3 )(\phi_2\phi_3)][\bar{\chi}_1 \phi_1] \rightarrow
[(\bar{\chi}_1  \bar{\chi}_2 \bar{\chi}_3 )\phi_2]+[(\phi_3 \phi_1)]
 ~.\label{eq:mynum}
\end{eqnarray}

The $W^\pm$ and $Z^0$ are also composite:
\begin{eqnarray}
&W^- = \frac{1}{\sqrt{3}}(\bar{\chi}_1  \bar{\chi}_2 \bar{\chi}_3 )
\{(\phi_2\phi_3)\phi_1+(\phi_3\phi_1)\phi_2+(\phi_1\phi_2)\phi_3\}~, \nonumber\\
&Z^0 \sim \{\bar{\phi}_1\phi_1+\bar{\phi}_2\phi_2+\bar{\phi}_3\phi_3\nonumber\\
&+(\phi_2\phi_3)(\bar{\phi}_2\bar{\phi}_3)
+(\phi_3\phi_1)(\bar{\phi}_3\bar{\phi}_1)
+(\phi_1\phi_2)(\bar{\phi}_1\bar{\phi}_2)\}~.\label{eq:mynum}
 \end{eqnarray}

The fermion mixing originates from flavor symmetry breaking. Look at the lepton mixing first. The notion of flavor is dynamical. For example, $\nu_e$ is the neutrino which is produced with $e^+$, or produces an $e^-$ in charged current weak interaction processes:
 \begin{eqnarray}
             & W^+\rightarrow l_\alpha^+ +\nu_\alpha,~~\alpha=e, \mu, \tau.
 \end{eqnarray}

If the symmetry $SU(3)_f\otimes U(1)_w$ was unbroken, $\bar{\phi}_1$, for example, could only interact with  $(\bar{\phi}_2\bar{\phi}_3)$
so that lepton numbers are conserved. However, if the flavor symmetry is broken by the condensate of two $\bar{\phi}$ preons, $\bar{\phi}_1$ can also interact with $(\bar{\phi}_3\bar{\phi}_1)$ and  $(\bar{\phi}_1\bar{\phi}_2)$, resulting in flavor mixing.

In this sense the three states given by (\ref{neutrino}) are neutrino mass eigenstates:
 \begin{eqnarray}
              \nu_1=(\bar{\phi}_2\bar{\phi}_3),~~
              \nu_2=(\bar{\phi}_3\bar{\phi}_1),~~
              \nu_3=(\bar{\phi}_1\bar{\phi}_2).
 \end{eqnarray}
The weak interaction eigenstates are superpositions of above three neutrinos $\nu_j(j=1,2,3)$:
 \begin{eqnarray}
              |\nu_\beta \rangle =\sum_\beta U_{\beta j} |\nu_j \rangle~~~\beta=e, \mu, \tau,
 \end{eqnarray}
where $U$ is a unitary matrix which is often called the Pontecorvo-Maki-Nakagawa-Sakata (PMNS) or Maki-Nakagawa-Sakata (MNS) mixing
matrix. \cite{Pontecorvo:1967fh, Pontecorvo:1957cp, Maki:1962mu}. The charged current of lepton weak interaction is as follows:
\begin{eqnarray}
  J_l^+=\begin{array}{ccc}
%                         & \begin{array}{ccc}  \nu_1  &  \nu_2  &  \nu_3   \end{array}  &  \\
                  \left(\begin{array}{ccc} e^+  & \mu^+ & \tau^+ \end{array}\right)
                  & \left( \begin{array}{ccc}
                  U_{e1} & U_{e2}  & U_{e3} \\
                   U_{\mu1} & U_{\mu2}  & U_{\mu3} \\
                   U_{\tau1} & U_{\tau2}  & U_{\tau3}\end{array}\right)
                   & \left(\begin{array}{c} \nu_1  \\ \nu_2 \\ \nu_3 \end{array}\right)
                     \end{array}.
\end{eqnarray}
      In terms of preons , the above expression takes the form:
\begin{eqnarray}
              J_l^+ =
                &U_{e1}(\chi_1  \chi_2 \chi_3 )\bar{\phi}_1(\bar{\phi}_2\bar{\phi}_3)+
                U_{\mu1}(\chi_1  \chi_2 \chi_3 )\bar{\phi}_2(\bar{\phi}_2\bar{\phi}_3)\nonumber\\
                &+U_{\tau1}(\chi_1  \chi_2 \chi_3 )\bar{\phi}_3(\bar{\phi}_2\bar{\phi}_3)
                +U_{e2}(\chi_1  \chi_2 \chi_3 )\bar{\phi}_1(\bar{\phi}_3\bar{\phi}_1)\nonumber\\
                &+U_{\mu2}(\chi_1  \chi_2 \chi_3 )\bar{\phi}_2(\bar{\phi}_3\bar{\phi}_1)+
                U_{\tau2}(\chi_1  \chi_2 \chi_3 )\bar{\phi}_3(\bar{\phi}_3\bar{\phi}_1)\nonumber\\
                &+U_{e3}(\chi_1  \chi_2 \chi_3 )\bar{\phi}_1(\bar{\phi}_1\bar{\phi}_2)+
                U_{\mu3}(\chi_1  \chi_2 \chi_3 )\bar{\phi}_2(\bar{\phi}_1\bar{\phi}_2)\nonumber\\
                &+U_{\tau3}(\chi_1  \chi_2 \chi_3 )\bar{\phi}_3(\bar{\phi}_1\bar{\phi}_2)
                ~.\label{eq:mynum}
\end{eqnarray}
The elements of lepton mixing matrix, $U_{\beta j}$, characterize the strength of interactions between $(\chi_1  \chi_2 \chi_3 )\bar{\phi}_1$ and $(\bar{\phi}_2\bar{\phi}_3)$, $\textit{etc.}$

The origin of quark flavor mixing is in common with  lepton mixing.
The d-type quarks given by (\ref{dquarks}) are  also mass eigenstates. They are
related with quark weak interaction eigenstates by the unitary quark flavor mixing matrix, CKM matrix $V$\cite{Cabibbo:1963yz, Kobayashi:1973fv}.
The charged current of quark weak interaction is given by
\begin{eqnarray}
           J_q^- =\bar{u}^i V_{ij}d^j,~~i=u, c, t; ~j=d, s, b~~.
\end{eqnarray}

In terms of preons , the quark charged current takes the form:
\begin{eqnarray}
J^- =
&\{V_{ud}(\bar{\chi}_2 \bar{\chi}_3 )(\phi_2\phi_3)\bar{\chi}_1  \phi_1+
  V_{us}(\bar{\chi}_2 \bar{\chi}_3 )(\phi_2\phi_3)\bar{\chi}_1 \phi_2\nonumber\\
  &+V_{ub}(\bar{\chi}_2 \bar{\chi}_3 )(\phi_2\phi_3)\bar{\chi}_1 \phi_3
+V_{cd}(\bar{\chi}_2 \bar{\chi}_3 )(\phi_3\phi_1)\bar{\chi}_1 \phi_1\nonumber\\
  &+V_{cs}(\bar{\chi}_2 \bar{\chi}_3 )(\phi_3\phi_1)\bar{\chi}_1 \phi_2+
  V_{cb}(\bar{\chi}_2 \bar{\chi}_3 )(\phi_3\phi_1)\bar{\chi}_1 \phi_3\nonumber\\
&+V_{td}(\bar{\chi}_2 \bar{\chi}_3 )(\phi_1\phi_2)\bar{\chi}_1 \phi_1+
  V_{ts}(\bar{\chi}_2 \bar{\chi}_3 )(\phi_1\phi_2)\bar{\chi}_1 \phi_2\nonumber\\
&+V_{tb}(\bar{\chi}_2 \bar{\chi}_3 )(\phi_1\phi_2)\bar{\chi}_1 \phi_3\}
~.\label{eq:mynum}
 \end{eqnarray}
 The elements of quark mixing matrix, $V_{ij}$, characterize the strength of interactions between $(\bar{\chi}_2 \bar{\chi}_3 )(\phi_2\phi_3)$ and $\bar{\chi}_1  \phi_1$, \textit{etc.}

\section{Summary and Discussions}

In this paper, we propose that there are six anticommutating fields called Hexad Preons. Quarks, leptons and W Z bosons are composite states with even number of preons and antipreons (complex conjugates of Hexad Preons). It is shown that there are exactly six quarks and six leptons with evident three families (generations) as well as a possible dark matter candidate. The underlying preon dynamics is supposed to be $U(3)\otimes U(3)$ local gauge theory which is identified with  $U(1)_Q\otimes SU(3)_C\otimes SU(3)_f\otimes U(1)_w$.  All quantum numbers of leptons and quarks can be given out of that of preons. Electro-strong interaction $U(1)_Q\otimes SU(3)_C$ gives the same dynamics as in ordinary QED and QCD. Weak interactions has no link to electromagnetic interaction so that there is no question of electroweak symmetry breaking. However, the symmetry of weak sector, $SU(3)_f\otimes U(1)_w$, is broken by the condensate of pairs of preons. The weak interaction is just residual "Van der Waals" forces between preons and dipreons. Both quark flavor mixing and lepton mixing are manifestation of this symmetry breaking.

Our present discussion is limited to the qualitative consequences of the model. More quantitative analysis, especially on the dynamics aspects of the model, is obviously needed. In the following, we will give some comments on three related problems.

The first problem is the difference between the electro-strong gauge interaction and that of the weak sector.
The electro-strong interaction among $\chi$ preons seems to be more "democratic", while the flavor gauge
interaction among $\phi$ preons prefer to bind two preons into dipreons. Since they both transform under
the gauge group $U(3)$, this problem must be related to the origin of the $U(3)\otimes U(3)$ gauge group.
In fact, the author has suggested that at very high energy Hexad Preons may carry hypercolor degree of
freedom transforming under $U(3,3)$ gauge group\cite{Wang:2010bt}. After the emergency of metric, the gauge group $U(3, 3)$
is broken down to its maximal compact subgroup $U(3)\otimes U(3)$. This means that the gravity may be
viewed as a symmetry-breaking effect in quantum field theory.
%The difference between the two $U(3)$ gauge interaction  The subsequent condensate of Hexad Preons,
%now transforming under $U(3)\otimes U(3)$, may form leptons, quarks, dark matter, \textit{etc.} as described above.
%This means that the gravity, electromagnetism, weak, and strong interaction may have the same origin.
So the careful examination of the mechanism for the $U(3, 3)$ hypercolor gauge symmetry breaking will not only give the clues to the difference between the two $U(3)$ gauge symmetries, but also help understanding the nature of gravity.

The second problem is the mechanism for $SU(3)_f\otimes U(1)_w$ gauge symmetry breaking. The perfect resolution of this problem should be able to recover all phenomenologically successful aspects of the standard electroweak theory. By the way, the mechanism for fermion mixing will also be elucidated.

The last question is on the role of $U(1)_w$ symmetry. The problem of fermion masses and mixing have been one of the outstanding puzzles in the standard model. One promising approach is the idea of family symmetry,
in particular the idea of a U(1) family symmetry\cite{Froggatt:1978nt}. In this respect, $U(1)_w$ symmetry may play a important role here. In addition, it has been shown that models which satisfy the Gatto-Sartori-Tonin relations \cite{Gatto:1968ss} must have both positive and negative abelian charges\cite{Fritzsch:1999ee}. In our model, quarks have both $U(1)_Q$ electric charge  and $U(1)_w$ weak charge with opposite sign. Therefore the relation between $U(1)_w$ symmetry and the fermion mass generation, fermion mixing needs further investigation.

\begin{acknowledgments}
This work is supported by Shanghai University of Engineering Science under grant No.11XK11 and No.2011X34.
\end{acknowledgments}

%%%%%%%%%%%%%%%%%%%%%%%%%%%%%%%%%%%%%%%%%%%%%%%%%%%%%%%%%%%%%%%%%%%%%%%%%%%%%

%%%%%%%%%%%%%%%%%%%%%%%%%%%%%%%%%%%%%

\end{document}